\documentstyle[12pt]{article}

\newcommand{\be}{\begin{equation}}
\newcommand{\ee}{\end{equation}}
\newcommand{\bea}{\begin{eqnarray}}
\newcommand{\eea}{\end{eqnarray}}

\begin{document}
\begin{titlepage}

\vspace{1in}

\begin{center}
\Large
{\bf On the  Cosmology and Symmetry of Dilaton--Axion Gravity}

\vspace{1in}

\normalsize

\large{James E. Lidsey$^1$}

\normalsize
\vspace{.7in}

{\em Astronomy Unit, School of Mathematical 
Sciences,  \\ 
Queen Mary \& Westfield, Mile End Road, LONDON, E1 4NS, U.K.}

\end{center}

\vspace{1in}

\baselineskip=24pt
\begin{abstract}
\noindent

A global O$(2,2)$ symmetry is found in the Brans--Dicke theory of
gravity when the dilaton is coupled to axion and moduli fields.  The
symmetry is broken if a cosmological constant is introduced. Within
the class of spatially homogeneous Bianchi cosmologies, only the type
I and V models respect the symmetry.  Isotropic cosmological solutions
are found for arbitrary spatial curvature. In the region of parameter
space relevant to the pre--big bang scenario, the interplay between
the scalar fields results in a bouncing cosmology. 
\end{abstract}


\vspace{.7in}
$^1$Electronic mail: jel@maths.qmw.ac.uk
 
\end{titlepage}


\setcounter{equation}{0}

In the standard inflationary scenario, the early
Universe becomes temporarily dominated by the self--interaction
potential energy of a scalar field \cite{inflation}.  This
potential plays the role of an effective cosmological constant 
$\Lambda$ and results in a quasi--de Sitter expansion of
the Universe. The energy stored in the false vacuum must exceed the
electroweak scale if baryogenesis is to proceed after
inflation has ended. However, 
it is well established from astrophysical observations that the
current value of the cosmological constant is extremely
small and corresponds to an energy of only $\le 0 .003$ eV 
\cite{currentvalue}. This is
approximately 121 orders of magnitude smaller than the Planck scale.  
A problem therefore arises when one attempts to
reconcile a relatively large vacuum energy in the early Universe with
the small value inferred today.

Recently, an alternative inflationary picture known as `pre--big
bang' cosmology  has been developed within the context of string
theory \cite{prebigbang}. In this scenario the accelerated expansion
is driven by the {\em kinetic} energy of the dilaton field rather than
its potential energy \cite{levin}. 
In principle, therefore, inflation may proceed without
the need for a cosmological constant. The Universe evolves from
the string perturbative vacuum into a regime of
high curvature and strong coupling.  The O$(3,3)$ invariance 
(T--duality) of the four--dimensional 
effective action relates this inflationary phase
of the Universe's history to a corresponding decelerating branch
\cite{tduality,ms}.  It is not currently known how a graceful exit from
inflation might proceed, however, because the two branches are
separated by a curvature singularity \cite{gracefulexit}.

Another symmetry of string theory that has received considerable
attention recently is S--duality \cite{sduality}.  This symmetry
relates the strong and weak coupling regimes of the theory.  Kar,
Maharana and Singh have addressed the cosmological constant problem
within the context of this symmetry by invoking 't Hooft's
`naturalness' hypothesis \cite{kar,hooft}.  In short, this hypothesis
states that it is natural for a physical parameter to be small if the
symmetry of the system is enhanced when that parameter vanishes. Thus,
the electron mass $m_e$ is a naturally small parameter because
specifying $m_e =0$ results in an additional chiral symmetry in the
action.  The field equations of the four--dimensional string effective
action are invariant under S--duality if and only if $\Lambda =0$
\cite{kar}.  Consequently, the symmetry of the theory is enhanced by
specifying $\Lambda =0$ and a small value for $\Lambda$ may be viewed
as natural, in the sense of 't Hooft \cite{hooft}.  It is possible,
therefore, that S--duality determines the value of the cosmological
constant and that T--duality leads to the inflationary expansion of
the Universe.

The coupling between the dilaton and graviton in the string effective
action arises in the Brans--Dicke theory of gravity \cite{bd},
where the coupling constant takes the specific value $\omega =-1$. In
view of the above developments, it is interesting to consider the
symmetric nature of the Brans--Dicke theory for $\omega \ne -1$. In
this paper, we find that the theory exhibits a global O$(2,2)$
invariance if the dilaton is appropriately coupled to other scalar
fields. The symmetry is broken for non--zero $\Lambda$.  We then
derive the cosmological solutions of this theory and relate them to
the pre--big bang scenario. 

We begin with an action of the form 
\be
\label{startingaction}
S=\int d^4 x \sqrt{-g}
e^{-\Phi} \left[ R- \omega \left( \nabla \Phi \right)^2 -
\frac{1}{2}  \left( \nabla \beta \right)^2 -\frac{1}{2} e^{(\gamma +1) \Phi} 
\left( \nabla \sigma \right)^2  -2\Lambda  \right]  ,
\ee
where $\Phi$ represents the dilaton field, $\beta$ is a `modulus'
field, $\sigma$ may be viewed as an axion--type field and $\Lambda$ is
assumed to be constant. The metric has signature $(-,+,+,+)$. The
constants $\omega$ and $\gamma$ determine the dilaton--graviton and
axion--dilaton couplings, respectively, and we assume $\omega > -3/2$
and $\gamma \ne -1$.  Theory (\ref{startingaction}) is well motivated
from a cosmological point of view. The modulus field may arise through
the spontaneous compactification of higher dimensions and the axion
field plays a dual role to that of the antisymmetric tensor potential
$B_{\mu\nu}$. 

For the spatially flat Friedmann--Robertson--Walker (FRW) Universe 
with line element $ds^2=-dt^2 +e^{2\alpha (t)} d{\bf x}^2$
and scale factor $a=e^{\alpha}$, 
the vacuum limit $(\sigma =\beta =0)$ of theory (\ref{startingaction}) 
is invariant under the discrete scale factor duality
\bea
\label{sfd}
\bar{\alpha} =\left( \frac{2+3\omega}{4+3\omega} \right) 
\alpha - 2\left( \frac{1+\omega}{4+3\omega} \right) \Phi \nonumber \\
\bar{\Phi} = -\left( \frac{6}{4+3\omega} \right) \alpha 
- \left( \frac{2+3\omega}{4+3\omega} \right) \Phi
\eea
if $\omega \ne -4/3$ \cite{lidsey}. 
This symmetry is an extension of the scale factor
duality of the string effective action. 
The cosmological solutions are given by \cite{hanlon}
\be
\label{asymptotic}
a^{(\pm )} \propto t^{p_{\pm}} ,\qquad 
e^{\Phi} \propto t^{3p_{\pm} -1}
\ee
for vanishing $\Lambda$, where 
\be
p_{\pm} \equiv \frac{1}{4+3\omega} \left[ 1+\omega \pm 
\left( 1+\frac{2\omega}{3} \right)^{1/2} \right]   .
\ee
The $(+)$-- and $(-)$-- branches are related by the scale
factor duality (\ref{sfd}). When $-4/3 < \omega < 0$, the
time--reversal of the $(-)$--branch corresponds to an accelerated
expansion, whereas the $(+)$--branch represents a decelerating
Universe.  These solutions form the basis of the pre--big bang
scenario and the two branches are separated by singularities in the
curvature and effective coupling at $t=0$.

The terms in Eq. (\ref{startingaction}) containing the cosmological
constant and modulus field are proportional to the `shifted' dilaton
field $\psi \equiv 3 \alpha -\Phi$ and this is invariant under the
transformation (\ref{sfd}), i.e., $\bar{\psi} =\psi$.  Thus, the
duality is respected for non--vanishing $\Lambda$ and $\beta$, but it
is broken by the axion field. On the other hand, this field leads 
to a global symmetry of the theory that becomes apparent in the
Einstein frame.  We therefore proceed by rescaling the metric such
that $\tilde{g}_{\mu\nu} =\Omega^2 g_{\mu\nu}$, where $\Omega^2
\equiv e^{-\Phi}$, and redefining the dilaton field $\varphi \equiv
(3+2\omega)^{1/2}
\Phi$. It follows that action (\ref{startingaction}) transforms
to\footnote{We drop tildes in what follows for notational
simplicity.}
\be
\label{einsteinaction}
S=\int d^4x \sqrt{-\tilde{g}} 
\left[ \tilde{R} -\frac{1}{2} \left( \tilde{\nabla} \varphi \right)^2 
-\frac{1}{2}
\left( \tilde{\nabla} \beta \right)^2 -\frac{1}{2} e^{-2\lambda \varphi}
\left( \tilde{\nabla} \sigma \right)^2 
-2\Lambda e^{\varphi/(3+2\omega )^{1/2}} \right]   ,
\ee
where $2\lambda  \equiv - (\gamma +1)/(3+2\omega )^{1/2}$. 
The string effective action corresponds 
to $\omega =-1$ and $\gamma =1$. We 
then define the symmetric $4 \times 4$ matrix \cite{ms}
\be
\label{M}
M=  \left( \begin{array}{cc} G^{-1} &  -G^{-1}B \\ 
BG^{-1} & G-BG^{-1}B \end{array} \right)   ,
\ee
where $G$ and $B$ are the $2 \times 2$ matrices
\be
\label{G+B}
G = \left( \begin{array}{cc} e^{\lambda (\varphi + \beta )} & 
0 \\ 0 & e^{\lambda (\varphi -\beta)} \end{array} \right) ,
\qquad 
B = \left( \begin{array}{cc} 0 & \lambda \sigma  \\ -
\lambda \sigma & 0 \end{array} \right)   .
\ee
The kinetic sector of action (\ref{einsteinaction}) may be 
written in the form
\be
\left( \nabla \varphi \right)^2 + \left( \nabla \beta \right)^2 
+ e^{-2\lambda \varphi} \left( \nabla \sigma \right)^2
 = -\frac{1}{2\lambda^2} {\rm tr} \left[ \left( 
G^{-1} \nabla B \right)^2 -\left( G^{-1} \nabla G \right)^2 \right]
\ee
and this implies that 
\be
\label{Maction}
S=\int d^4 x \sqrt{-g} \left[ R+ \frac{1}{8\lambda^2} {\rm tr} 
\left( \nabla M \nabla M^{-1} \right) 
-2\Lambda e^{\varphi/(3+2\omega )^{1/2}} \right]    .
\ee

The matrix (\ref{M}) satisfies the constraint 
$M^T \eta M =\eta$, where 
\be
\label{eta}
\eta \equiv \left( \begin{array}{cc} 0 & {\rm I}_2 \\ {\rm I}_2 &  0 
\end{array} \right)
\ee
is an O$(2,2)$ metric in non--diagonal form 
and ${\rm I}_2$ is the identity matrix 
in 2 dimensions. Eq. (\ref{M}) 
is therefore an element of the group O$(2,2)$. 
Moreover, it can be verified that action (\ref{einsteinaction}) is invariant 
under the global O$(2,2)$ transformation 
\be
\label{O22}
\bar{M} =  \Sigma M \Sigma^T, \qquad \bar{g}_{\mu\nu}=g_{\mu\nu} ,
\qquad \Sigma^T \eta \Sigma =\eta
\ee
if the cosmological constant vanishes. The symmetry 
is broken if the cosmological constant is non--zero 
because the 
dilaton is not invariant under the action of Eq. (\ref{O22}). 
The symmetry 
of the theory is therefore enhanced if $\Lambda =0$, so 
the vanishing of $\Lambda$ is natural in this 
theory in the sense advocated by  't Hooft \cite{hooft}. 
We remark that if $\Lambda$ is in general  
some function of the dilaton field, the symmetry (\ref{O22}) 
is only respected for the specific form $\Lambda 
\propto e^{-\varphi/(3+2\omega )^{1/2}}$. 

The symmetry (\ref{O22}) is related to 
the global SL$(2,R)$ symmetry associated with the  
four--dimensional SL$(2,R)/U(1)$ non--linear $\sigma$--model 
\cite{sigma}. This
becomes apparent after the introduction of the complex
fields \cite{compare}
\be
\label{TU}
T \equiv T_1+iT_2 = \lambda  \sigma +ie^{\lambda \varphi}, \qquad 
U \equiv ie^{\lambda \beta}
\ee
so that Eq. (\ref{M}) takes the form 
\be
M=\frac{i}{T_2U} 
 \left( \begin{array}{cccc} 1 & 0 & 0 & -T_1 \\
0 & |U|^2 & T_1|U|^2 & 0 \\
0 & T_1 |U|^2 & |T|^2|U|^2 & 0 \\
-T_1 & 0 & 0 & |T|^2  
\end{array} \right)  .
\ee
We may now consider the specific element of 
O$(2,2)$ given by \cite{sabra} 
\be
\Sigma = \left( \begin{array}{cccc} d & 0 & 0 & -c \\
0 & d & c & 0 \\
0 & b & a & 0 \\
-b & 0 & 0 & a 
\end{array} \right)    ,
\ee
where $ad-bc= 1$. Eq. (\ref{O22}) then generates the 
SL$(2,R)$ transformation
\be
\label{SL2RT}
\bar{T} =\frac{aT+b}{cT+d}, \quad \bar{U}=U    . 
\ee
This leaves the modulus invariant, but combines the dilaton and axion 
in a non--trivial way. 

An alternative formulation of action (\ref{Maction}) is possible when 
$\Lambda =0$. We define the matrix \cite{gk}
\be
\label{N}
N\equiv \left( \begin{array}{cc} P^{-1} & P^{-1} Q \\
Q P^{-1} & P+QP^{-1} Q  \end{array} \right)   ,
\ee
where $P$ and $Q$ are the $2 \times 2$ matrices
\be
P =  \left( \begin{array}{cc} e^{\lambda \varphi} & 0 \\
0 & e^{\lambda \beta} \end{array} \right) , 
\qquad 
Q =  \left( \begin{array}{cc} \lambda \sigma & 0 \\ 0 & 0 
 \end{array} \right)    .
\ee
Eq. (\ref{N}) satisfies the constraint $N^T JN=J$, where 
\be
\label{J}
J \equiv \left( \begin{array}{cc} 0 & {\rm I}_2 \\ -{\rm I}_2 & 0 
\end{array} \right)
\ee
and it is therefore an element of the real symplectic group Sp$(4,R)$.
Moreover, the inverse of $N$ is given  by 
$N^{-1} =-J N^T J$, since $J^2 = -{\rm I}_2$. Thus, the 
action is given by 
\be
\label{Naction}
S=\int d^4 x \sqrt{-g} \left[ R+ \frac{1}{4\lambda^2} {\rm tr} 
\left( \nabla N \nabla N^{-1} \right) \right]
\ee
and the theory is invariant under the global Sp$(4,R)$ transformation
\be
\label{sp4r}
\bar{N} = \Theta N \Theta^T , \qquad \bar{g}_{\mu\nu} =g_{\mu\nu} , \qquad 
\Theta^T J \Theta = J    .
\ee
The effect of Eq. (\ref{sp4r}) on the complex symmetric matrix $Z
\equiv Q +iP = {\rm diag} [T,U]$ 
is analogous to that of the SL$(2, R)$ transformation
(\ref{SL2RT}) on the complex dilaton--axion field $T$ \cite{gk}. In a
sense, therefore, Eq. (\ref{sp4r}) may be viewed as a
matrix--valued SL$(2,R)$ transformation.

The global symmetry associated with action (\ref{startingaction})
provides motivation for considering this theory further within a
cosmological context. The class of spatially homogeneous cosmologies
admits a Lie group of isometries that act transitively on the
space--like three--dimensional orbits \cite{wald}. 
The anisotropy of each model 
is determined by the structure constants of the Lie algebra of
$G_3$ and the metric on the three--surfaces 
may be written as 
\be
h_{ab} (t) = e^{2 \alpha
(t)}\left( e^{2\beta (t)} \right)_{ab} , \qquad a,b = 1,2,3   ,
\ee
where
$\beta_{ab} \equiv {\rm diag} \left[ \beta_+
+\sqrt{3} \beta_-,
\beta_+ -\sqrt{3} \beta_- ,-2 \beta_+ \right]$ is a traceless matrix
and $e^{3\alpha}$ represents the effective spatial volume
of the Universe. 

We only consider those models for which a Lagrangian formulation 
is possible \cite{bianchilang}. 
Integration over the spatial variables in action
(\ref{einsteinaction}) then implies that
\be
\label{bianchiaction}
S=\int dt e^{3\alpha} \left[ -6\dot{\alpha}^2 +6\dot{\beta}^2_+ + 
6\dot{\beta}^2_- +\frac{1}{2} \dot{\varphi}^2 + \frac{1}{2} e^{-2
\lambda \varphi} \dot{\sigma}^2 + e^{-2\alpha} U(\beta_{\pm}) \right]
\ee
for $\beta = \Lambda = 0$, where a dot denotes differentiation with
respect to time, a boundary term has been neglected and the comoving
volume of the Universe has been normalized to unity.  The function
$U(\beta_{\pm})$ determines the scalar curvature of the
three--surfaces and is different for each Bianchi type
\cite{wald1}.

The fields $\beta_{\pm}$ play the role of moduli and formally we
may identify $d\beta^2 =12 (d\beta_+^2 + d\beta_-^2)$.  After
comparing Eqs. (\ref{einsteinaction}) and (\ref{bianchiaction}), we
then deduce that the kinetic sector of Eq. (\ref{bianchiaction})
is symmetric under the global O$(2,2)$ transformation 
\be
\label{o22bianchi}
\bar{M}=\Sigma M \Sigma^T , \qquad \bar{\alpha} =\alpha , \qquad
\Sigma^T \eta \Sigma =\eta   .
\ee
This transformation alters the degree of anisotropy in 
each of the three spatial directions whilst  preserving the spatial 
volume of the Universe. 
However, Eq. (\ref{bianchiaction}) also contains an effective
potential for the moduli.  Since Eq. (\ref{o22bianchi}) relates the
fields in a non--linear fashion, this term is only invariant under
such a transformation when it is {\em independent} of $\beta_{\pm}$.
Thus, $U$ must be constant if the full action (\ref{bianchiaction}) is
to be symmetric and this condition is only satisfied for the Bianchi
types I and V \cite{bianchilang,wald1}. These models represent the
anisotropic generalizations of the spatially flat and negatively
curved FRW Universes, respectively. Within this context, therefore,
the type I and V models are more symmetric than the other Bianchi
types and it is interesting that they are both associated with
isotropic FRW Universes.

We now proceed to solve the field equations for the class of FRW Universes. 
In general, the line element of these models is 
\be
\label{FRWmetric}
ds^2 =-dt^2 + e^{2\alpha (t)} \left[ \frac{dr^2}{1-kr^2} +r^2 
\left( d\theta^2 +{\rm sin}^2 \theta d\psi^2 \right) \right]   ,
\ee
where the curvature parameter takes values $-1, 0, +1$ for negatively
curved, flat and positively curved models, respectively. It 
proves convenient to express the field
equations derived from action (\ref{startingaction})
in terms of the conformally invariant time parameter
$\eta \equiv \int dt e^{-\alpha (t)}$, the Lorenz--Petzold
variable $X \equiv e^{2\alpha -\Phi}$ \cite{LP} and 
the rescaled dilaton field
$Y \equiv -(1+2\omega /3)^{1/2} \Phi$. The action  (\ref{startingaction})
with $\Lambda =0$ then reduces to
\be
S=\int d\eta X \left[ -\frac{3}{2} \frac{X'^2}{X^2}
+\frac{3}{2} Y'^2 +\frac{1}{2} \beta'^2  +
\frac{1}{2} e^{-\epsilon Y} \sigma'^2  +6k \right]   ,
\ee
where a prime denotes differentiation with respect to $\eta$ and 
$\epsilon \equiv (1+\gamma )/(1+2\omega /3)^{1/2}$. 

The cosmological field equations are 
given by 
\bea
\label{cosmo1}
X \beta '  =p , \qquad Xe^{-\epsilon Y} \sigma ' =q \\
\label{cosmo3}
X'' +4kX =0 \\
\label{cosmo4}
\left( XY' \right) ' =-\frac{\epsilon q^2}{6X}  e^{\epsilon Y} \\
\label{cosmo5}
3 \frac{X'^2}{X^2} -3Y'^2 +12 k -\beta'^2 -e^{-\epsilon Y} \sigma'^2 =0  ,
\eea
where $p$ and $q$ are 
arbitrary constants. The advantage of employing the Lorenz--Petzold
variable is that Eq. (\ref{cosmo3}) is identical to the corresponding  
expression for the vacuum model $(\beta = \sigma =0)$ 
\cite{mw}. Its 
first integral may be written as 
\be
\label{firstintegral}
X'^2+4kX^2 =A^2+p^2   ,
\ee
where $A$ is an arbitrary constant. It may then be verified 
by direct substitution of Eq. (\ref{firstintegral}) into Eq. 
(\ref{cosmo5}) that the latter equation is a first integral
of Eq. (\ref{cosmo4}). Thus, 
the evolution of the scale factor may be determined by solving 
the first--order equations (\ref{cosmo5})
and (\ref{firstintegral}). 

The solution to Eq. (\ref{firstintegral}) is given by
\be
\label{Xsolution}
X(\tau) = \sqrt{A^2+p^2} \frac{\tau}{1+k\tau^2}    ,
\ee
where we have defined \cite{mw}
\bea
\label{newtime}
\tau =  \left\{ \begin{array}{cc} \eta , & k =0 \\ {\rm tan} \eta , 
& k=+1 \\ {\rm tanh}  \eta , & k=-1 \end{array} \right.   . 
\eea
Substitution of Eq. (\ref{Xsolution})
into Eq. (\ref{cosmo5}) then leads to the solution
\bea
\label{dilatonsolution}
e^{ (1+\gamma )\Phi} = \frac{q^2}{12 \tilde{A}^2} \left[ 
\left( \frac{\tau}{\tau_0} \right)^n +\left( \frac{\tau}{\tau_0} 
\right)^{-n} \right]^{2} \\ \nonumber 
a^2 =(A^2+p^2)^{1/2} \left( \frac{q^2}{12 \tilde{A}^2} 
\right)^{1/(1+\gamma )} \left[ \left( \frac{\tau}{\tau_0} \right)^n 
+ \left(  \frac{\tau}{\tau_0} \right)^{-n} \right]^{2/(1+\gamma )} 
\frac{\tau}{1+k\tau^2}    ,
\eea
where $\tilde{A}^2 = A^2 +2p^2/3$, $\tau_0$ is an integration 
constant  and 
\be
\label{n}
n \equiv \frac{1}{2} \frac{1+\gamma}{(1+2\omega /3 )^{1/2}}
\left( \frac{A^2+2p^2/3}{A^2+p^2} \right)^{1/2}    .
\ee
This solution generalizes the stiff perfect fluid $(\beta =
\gamma =0)$ and string models  \cite{mw,fluid,copeland}. 

We will consider 
the evolution of the flat $(k=0)$ model. 
A necessary and sufficient condition for inflation is that the second
time derivative of the scale factor be positive definite, i.e.,
$\alpha '' >0$. When $p=0$, this requires  $\omega
<0$ and, indeed, the region of parameter space 
corresponding to $-4/3 <
\omega <0$ is relevant to the pre--big bang scenario. Although this
may appear a somewhat restrictive regime, it should be emphasized that
it includes the truncated string effective action and higher
dimensional Einstein gravity. If the latter theory is dimensionally
reduced to four dimensions with an isotropic, Ricci--flat internal
space of radius $b$ and dimensionality $d$, the resultant action is
given by the dilaton--graviton sector of Eq. (\ref{startingaction})
with $\Phi \propto -d \ln b$ and $\omega =-1 +1/d$. In view of this,
we will consider this range of $\omega$.

The qualitative behaviour of solution 
(\ref{dilatonsolution}) is not affected by the
moduli and we may therefore specify $p=0$ without loss of
generality. The sign of $n$ is important, however. We begin by
considering the case $n>0$ $(\gamma >-1)$. In the limit $t \rightarrow
0$, the asymptotic form of the solution is given by the $(-)$--branch
of the vacuum solution (\ref{asymptotic}). On the other hand, the late time 
limit $(t
\rightarrow \infty)$ is given by the
$(+)$--branch. Thus, the scale factor is initially infinitely large
and the Universe undergoes an accelerated contraction 
to a minimum size before reexpanding to
infinity. Inflationary behaviour is inevitable immediately 
after the bounce since the Universe must accelerate away from the 
point of maximum contraction. Inflation does not last
indefinitely, however, since $p_+ < 1$ when $-4/3 < \omega < 0$.  The
effective gravitational coupling $G_{\rm eff} \propto e^{\Phi}$ is
also bounded from below.  It is initially divergent and decreases as
the Universe expands to a minimum value of $G_{\rm eff}
=(q^2/3A^2)^{1/(1+\gamma )}$ at a time $\tau_0$.  It then increases
indefinitely for $\eta > \tau_0$.  The bound on the coupling is 
determined by the canonical momentum of the axion field. 

\vspace{.1in}
\centerline{\bf Figure 1} 
\vspace{.1in}

The evolution of the scale factor and 
effective coupling is shown in Figure 1 and is
comparable  to that found in the specific models of Refs. 
\cite{copeland,af}.  Our analysis shows that such behaviour does 
not depend too sensitively
on the coupling between the dilaton and axion fields and it may
therefore be quite generic in theories of this type.  
The curvature invariants diverge at $t=0$ and
there is an infinite proper distance between two given points in
space--time.  Thus, the Universe is singular at this point even though
it has infinite size.  
A singularity of this type has been termed an `anti--big bang'
singularity \cite{af}, since it is in contrast to the
conventional big bang singularity where the proper distance between
two points is zero.

The high and low curvature regimes are related by the scale
factor duality (\ref{sfd}) and the vacuum solutions 
are recovered in these limits because the kinetic
energy of the dilaton dominates the dynamics. The vacuum
Universe (\ref{asymptotic}) undergoes monotonic expansion or
contraction and the role of the axion is to induce a bounce between
the contracting and expanding branches 
\cite{copeland}. Thus, a classical transition
between the two branches is possible when an axion field is
present. 

For completeness, we remark that the role of the vacuum branches
(\ref{asymptotic}) is interchanged when $n<0$ $(\gamma < -1 )$. The
$(+)$--branch applies as $t \rightarrow 0$ and the $(-)$--branch as $t
\rightarrow  \infty$. Thus, the Universe has zero spatial volume 
initially and expands out of the singularity. There is now an upper
bound on the maximum size attained by the Universe, however, and it
recollapses after a finite proper time.  Although this solution does
not appear to be physically relevant, it would be interesting to
investigate whether perfect fluid matter sources can prevent the
recollapse from proceeding.  Similar qualitative behaviour is expected
in the vicinity of the singularity since the energy density of the
dilaton field dominates ordinary matter at early times
\cite{fluid}. However, the Universe would become dominated by the
fluid sources at later times.

In conclusion, we have identified a global O$(2,2)$ symmetry in the
Brans--Dicke theory of gravity that exists when the dilaton is coupled
to moduli and axion fields.  The symmetry is an extension to the
Brans--Dicke theory of the SL$(2,R)$ invariance of the truncated string
effective action. The symmetry is not respected when a cosmological
constant is present and a vanishingly small $\Lambda$ in this theory
is therefore consistent with 't Hooft's naturalness hypothesis.  When
the modulus field is related to the anisotropy parameters of the
spatially homogeneous Universes, the symmetry is only respected for
the Bianchi types I and V.  Cosmological solutions were found for
arbitrary spatial curvature and the flat model was discussed within
the context of the pre--big bang scenario.  The axion field induces a
classical transition between accelerating and decelerating phases by
causing the Universe to bounce.

\vspace{.7in}
{\bf Acknowledgments} The author is supported by the Particle Physics
and Astronomy Research Council (PPARC), UK.  We thank D. Wands for helpful 
discussions.

\newpage

\centerline{\bf Figure Caption}

\vspace{.2in}

{\em Figure 1}: (a) At early times the dilaton dominates the axion and
the Universe undergoes an accelerated contraction from an initially
singular state. The axion induces a bounce and the Universe reexpands
after a finite proper time into a decelerating phase. The high and low
curvature limits of the solution are determined by the 
$(-)$-- and $(+)$--branches of the vacuum solution and are 
related by the scale factor duality of the theory. (b) The 
effective gravitational coupling is initially divergent, but becomes 
progressively weaker until a lower bound is attained. The bound 
is determined by the momentum associated with the axion. The 
coupling then increases monotonically with time.

\end{document}